# Improved Segmentation and Detection Sensitivity of Diffusion-Weighted Brain Infarct Lesions with Synthetically Enhanced Deep Learning

Christian Federau[1], MD, MSc, Soren Christensen[3], PhD, Nino Scherrer[1], BSc, Johanna Ospel[2], MD, Victor Schulze-Zachau[2], MD, Noemi Schmidt[2], MD, Hanns-Christian Breit[2], MD, MSc, Julian Maclaren[3], PhD, Maarten Lansberg[3], MD, PhD, Sebastian Kozerke[1], PhD

[1]Institute for Biomedical Engineering, ETH Zürich und University of Zürich, Gloriastrasse 35  8092 Zürich, Switzerland
[2]Division of Diagnostic and Interventional Neuroradiology, Department of Radiology, University Hospital Basel, Petersgraben, 4031 Basel, Switzerland
[3]Stanford Stroke Center, Department of Neurology, Stanford University

*Corresponding author:
Christian Federau, Institute for Biomedical Engineering, ETH Zürich and University of Zürich, Gloriastrasse 35, 8092 Zürich, Switzerland, federau@biomed.ee.ethz.ch

**Abstract:**

**Purpose**

To compare the segmentation and detection performance of a deep learning model trained on a database of human-labelled clinical diffusion-weighted (DW) stroke lesions to a model trained on the same database enhanced with synthetic DW stroke lesions.

**Methods**

In this institutional review board approved study, a stroke database of 962 cases (average age 65±17 SD years, 255 males, 449 scans with DW positive stroke lesions) and a normal database of 2,027 patients (average age 38±24 SD years, 1088 females) were obtained. Brain volumes with synthetic DW stroke lesions were produced by warping the relative signal increase of real strokes to normal brain volumes. A generic 3D U-Net was trained on four different databases to generate four different models: *(a)* 375 neuroradiologist-labeled clinical DW positive stroke cases (CDB); *(b)* 2,000 synthetic cases (S2DB); *(c)* CDB+2,000 synthetic cases (CS2DB); or *(d)* CDB+40,000 synthetic cases (CS40DB). The models were tested on 20% (n=192) of the cases of the stroke database, which were excluded from the training set. Segmentation accuracy was characterized using Dice score and lesion volume of the stroke segmentation, and statistical significance was tested using a paired, two-tailed, Student's t-test. Detection sensitivity and specificity was compared to three neuroradiologists.

**Results**

The performance of the 3D U-Net model trained on the CS40DB (mean Dice 0.72) was better than models trained on the CS2DB (Dice 0.70, $P$ <0.001) or the CDB (Dice 0.65, $P$ <0.001). The deep learning model was also more sensitive (91%[89%-93%]) than each of the three human readers (84%[81%-87%], 78%[75%-81%], and 79%[76%-82%]), but less specific (75%[72%-78%] vs for the three human readers (96%[94%-97%], 92%[90%-94%] and 89%[86%-91%]).

**Conclusion**

Deep learning training for segmentation and detection of DW stroke lesions was significantly improved by enhancing the training set with synthetic lesions.

**Introduction**

The fast and accurate detection of stroke on radiological images is of utmost importance, because clinical outcome is directly related to the time to treatment (1,2). However, the detection of stroke lesions can be challenging and mistakes in diagnosis can lead to delays in therapy with potential harmful consequences for the patient. Even large vessel occlusion, especially if sub-occlusive, can have a small infarct core that could be missed and errors in the stroke core volume measurement can lead to an erroneous clinical decision. Missing a small embolic stroke can lead to patient discharge without adequate clarification of its source (such as endocarditis, open foramen ovale, carotid plaques) and without adequate prophylactic therapy.

Machine learning and, in particular, deep learning methods have tremendous potential for application to medical images, for example for detection, segmentation, or diagnosis of pathological conditions (3–6). These methods could help circumvent mistakes due to human tiredness, limited attention spans, distractions, and limited time resources. However, progress in machine learning for medical image analysis is hampered by the small size of the databases available to train models. For example, the database of the Ischemic Stroke Lesion Segmentation Challenge (ISLES) 2015 (7) contains only 64 stroke cases. The ISLES database is small-scale compared to the ImageNet database, (8) which currently holds more than 14 million human-labelled images that have been widely used to train models to identify everyday objects. There is little hope for a substantial improvement in the number of available labeled medical images for training machine learning models in the near future, as collecting a large number of labeled images is obstructed by ethical and privacy issues, medico-legal aspects, patient protection, and various technical challenges. Further, the labelling of radiological images is tedious, requires a high level of expertise, is error prone, and can be very subjective (see **Suppl. Figs 1-2** for examples).

We hypothesized that a database of clinical stroke lesion images could be enhanced with synthetic images of stroke lesions produced from clinical images that do not otherwise contain an abnormality. These 'normal' images are available in large numbers and could provide the data necessary for appropriate machine learning. We hypothesized that the performance of a network trained on a synthetically enriched dataset would be better than a network trained on clinical data alone.

In this work, we propose an algorithm to produce realistic synthetic magnetic resonance DW volumes of stroke lesions by extracting the general features of a DW stroke lesion from clinical volumes and combining these features with normal DW volumes. Due to the combinatorial possibilities, such an algorithm allows the generation of a vast number of training datasets, and because the lesions are generated artificially, the ground truth lesion location is known exactly.

The objectives of this study were threefold: *(a)* to generate a large number of diffusion-weighted (DW) volumes with synthetic stroke lesions from a stroke database and a normal database; *(b)* to independently train a generic 3D U-Net (9) on training sets comprising clinical cases only, synthetic volumes only, or the combination of clinical and synthetic volumes; *(c)* to compare the diagnostic accuracy in lesion segmentation and detection of these networks on a separate test set comprising volumes of patients with clinical suspicion of stroke.

**Methods**

Institutional Review Board approval (from the Ethikkommission Nordwest- und Zentralschweiz) was obtained.

Databases: Two databases were produced: a stroke database and a normal database. For the *stroke database*, patients who presented with symptoms of acute stroke and who had an MR examination within a span of 24 hours after admission, were identified and images were downloaded from the radiological information system - picture archiving and communication system (RIS-PACS) of the University hospital Basel, using an in-house PACS crawler. Images were reviewed by an experienced neuroradiologist (CF, 10 years of experience) for evidence of non-ischemic stroke-related abnormalities and cases that included intracranial bleeds, tumors, large chronic strokes, metallic artefacts (for example from metallic scalp clips after craniectomy, external ventricular derivation or aneurysm clips) or any other significant pathologies, were excluded. For the *normal database*, normal brain MR images, as defined by a report with normal finding, as evaluated by NS and HCB (1 year of experience) were downloaded from the PACS.

Databases Characteristics:

962 cases were included in the stroke database: 449 scans with DW positive stroke lesions (200 left-sided strokes; 193 right-sided strokes, 56 bilateral strokes; patient age [mean ± standard deviation] 72 ± 14 y; 194 females; 255 males) and 513 without DW positive stroke lesions (patient age 65 ± 17 y, 258 females, 255 males). 106 scans were excluded. 2,027 cases were included in the normal database (mean patient age 38 ± 24 y, 1088 females, 939 males). Acquisition parameters can be found in **Suppl. Table 1**. The Dice coefficient between the segmentation by the human reader 1 and 2 was 0.766 ± 0.139 for the full stroke database (n=962) and 0.769 ± 0.135 for the test set (n=192).

Images Post-Processing: Diffusion-weighted volumes of both databases were anonymized and co-registered with an affine transformation to the standard Montreal Neurological Institute space (10) rotated in the anterior commissure – posterior, and resampled to a standard resolution of 128 x 128 x 40 voxels, using the Advanced Normalization Tools (ANTS, (11)).

Manual Lesion Segmentation: All DW lesions were then manually segmented by two neuroradiologists (JO and VSZ, 2 years of experience, below *as human reader 1 and 2),* using a custom-made Horos (12) plug-in. The stroke segmentations of human reader 1 were then reviewed and corrected if necessary by an experienced neuroradiologist (CF, 10 years of experience, below as *human reader 3,* used as reference standard). The neuroradiologists were free to adjust the contrast level. All images and binary masks were saved in the DICOM format. Only the labels of human *reader 3* were used for training and testing. The labels of *human reader 1, 2* were used to calculate a human inter-observer Dice coefficient for comparison with the networks' Dice coefficient (see following Quantitative Analysis section).

Diffusion-weighted MR Image Acquisition: Images were acquired at a single institution, at 1.5T (Avanto Fit) and 3T Skyra and Skyra Fit) (all Siemens Healthineers, Erlangen, Germany), at a b-value of 1000 s/mm$^2$. Acquisition parameters can be found in **Suppl. Table 1**.

Synthetic Stroke Image Generation Algorithm: Synthetic DW stroke volumes were produced as follows. A stroke volume from the training set of the stroke database (n=375, see following Datasets for Training and Testing section) and a normal scan from the normal database (n=2,027) were selected randomly (with replacement). The stroke volume was coregistered to the normal image using a deformable b-spline transformation and third-order spline resampling using ANTS. The voxel-wise signal increase inside the stroke area, relative to the mean of the signal intensity in the healthy hemisphere, thresholded at a relative signal increase of at least 8%, was fused to the signal of the brain parenchyma of the normal image by multiplying voxel-wise the relative signal increase with the local normal signal of the parenchyma (**Fig. 1 A**). An anatomical correction filter was applied to remove the synthetic lesions outside the brain parenchyma. The binary mask of the synthetic lesion was created from the resampled lesion mask by setting all values above 0 to 1. Using this technique, a database of 40,000 synthetic 3D stroke volumes with a matrix of 128 x 128 x 40 voxels were produced and saved in the DICOM format.

Datasets for Training and Testing: The clinical stroke cases were split randomly into a training set (80%), from which only the DW positive cases (n=375) were included and a final test set (20%) (n=192), from which both DW positive (n=74) and negative (n=118) cases were kept. The same generic 3D U-Net (see following Neural Network section) was trained from scratch on the following four distinct training datasets (**Fig. 1 A, orange boxes**): *(a)* the database (DB) of 375 human-labeled clinical stroke cases (CDB), *(b)* 2,000 synthetic cases (i.e. without clinical cases) (S2DB), *(c)* the database of human-labeled clinical stroke cases and 2,000 synthetic cases (CS2DB), and *(d)* the database of human-labeled clinical stroke cases and 40,000 synthetic cases (CS40DB).

Neural network: A 3D U-Net (9) was implemented in Tensorflow (13). The network had a depth of three layers. The highest layer had 64 feature maps, and the number of feature maps doubled with each downsampling step (**Fig. 1 B**). The loss function consisted of the sum of a cross-entropy term and a weighted Dice coefficient term (background / foreground, i.e., normal brain parenchyma / DW lesion). We used the Adam optimizer (14), with a learning rate decreasing exponentially after initial warm-up steps. Batch normalization was used. The training was stopped after 1000 epochs.

Data normalization and augmentation: DW volumes were cropped to a volume of 96 x 80 x 40 voxel resolution, and signal intensity was clipped at the 99.5% percentile and normalized by dividing by the maximum after clipping. Data were augmented during the training on the fly using randomly a left-right flip of the images, as well as a maximum of two of the following transformations along all three spatial dimensions, chosen randomly, with maximum ranges in brackets (proportion of field of view): translation [-0.02, 0.02], zoom [0.95, 1.05], stretch [0.95, 1.05] and shear [-0.03, 0.03].

Cross-validation: In a first step, before the main training step of this work, the dropout rate and the weights of the Dice coefficient loss hyperparameters were optimized with cross-validation. The 80% of clinical cases reserved for training, were separated in five buckets of equivalent size. The cases were split such that each bucket contained an identical number of stroke cases and non-stroke cases. In each bucket, an equivalent number of synthetic stroke cases, derived from the clinical cases in the bucket, were added. Training for cross-validation was performed cyclically using four buckets of clinical and synthetic strokes, and tested on the clinical cases of the fifth bucket (**Suppl. Fig. 3**). The dropout rate was optimized between 40 % and 50 %, and the weights of the Dice coefficient loss between {0.1, 0.9}, {0.05, 0.95}, {0.01, 0.99}, and {0.001, 0.999}. The best set of those hyperparameters was selected based on the averaged Dice curve as function of training epochs (**Suppl. Fig. 4**), and used for the subsequent analysis.

Quantitative Analysis: The performance of the 3D U-Net trained on the four training datasets was measured on the test set, using two metrics: *(a)* Dice score between model prediction and label segmentation and *(b)* lesion volume. To exclude the initial part of the learning which is not relevant, and the later part where overfitting might occur, we defined a trained range on which the quantitative analysis was performed, as the range starting at the approximate level of flattening of the Dice coefficient curve and completing 200 epochs later. The quantitative analysis includes mean, standard deviation, median, minimum-maximum range of the Dice coefficients, as well as the error in stroke volume.

To compare the lesion detection capabilities of the networks and human neuroradiologists, a detailed analysis of stroke lesion detection was performed in a random subset of 80 cases of the test set comparing the predictions obtained by all models at a single epoch (chosen at the median of their trained range) to the human labelling. Because the human labelling can also be error prone, the images were reassessed in light of the results given by the models by placing all predictions and human labelling next to each other and having them scored blindly by three neuroradiologists in consensus. Stroke lesion detection sensitivity, specificity, positive and negative predictive values were obtained accordingly.

Statistical Analysis: A paired, two-tailed, Student's t-test was performed (pairwise between all combinations of the four models) to reject the null hypothesis that two Dice coefficient as function of epoch number, obtained from training on two different datasets, were similar in the trained range. Significance level was set to $\alpha < 0.001$. False discovery rate was corrected for using the Benjamini–Yekutieli procedure (15). All computations were run on Excel and Matlab.

## Results

Cross-Validation

The cross-validation curves comparing the Dice coefficient as function of epoch number were very similar for the hyperparameters tested (i.e. dropout rate and Dice loss weights, **Suppl. Fig. 4**). The best set of hyperparameter loss weights {0.01, 0.99} and dropouts 40% was selected from the cross-validation study for further analysis.

Model Trained on the CS40 Database Demonstrated Optimal Segmentation and Stroke Detection

Using the best set of hyperparameter found from the preliminary cross-validation, four 3D U-Net models were trained on the four different databases (CDB, S2DB, CS2DB, and CS40DB). The models' performance, as evaluated with the Dice score on the test set as function of epoch number (before overfitting occurred), gradually increased in the following order: from the least optimal model trained on the CDB, S2DB, CS2DB, to the most optimal model trained on the CS40DB (**Fig. 2**). The trained range for CS40DB was selected as epoch 100 to 300, while the trained range for CDB, S2BD, and CS2DB was 250 to 450 epochs. The mean Dice coefficient was $0.618 \pm 0.077$, $0.653 \pm 0.036$, $0.695 \pm 0.027$, and $0.721 \pm 0.010$ for the model trained on the CDB, S2DB, CS2DB, and CS40DB, respectively (**Table 1,** all $P < 10^{-7}$). Note that the stability of the prediction, as measured with a decrease in the standard deviation, increased gradually in the same order. The error in stroke volume dropped rapidly to clinically negligible values below 4 mL for all models for almost all epochs above 300 epochs (**Suppl. Fig. 6**).

The detailed lesion detection analysis (**Table 2**), in which a reassessment of the human labelling was performed in light of the finding of the networks, demonstrated that the best deep learning model, trained on the CS40DB, was more sensitive (686/753 (91%, 95% confidence interval [89% - 93%])) than all three human readers (635/753 (84% [81% - 87%]), 586/753 (78% [75% - 81%]) and 596/753 (79% [76% - 82%])), and was particularly efficient at picking up subtle signal changes in neighboring slices of distinct lesions (**Fig. 3** and **Suppl. Fig. 7-8**). The three human readers were more specific than all the models (649/674 (96% [94% - 97%]), 621/674 (92% [90% - 94%]) and 599/674 (89% [86% - 91%]) versus 506/674 (75% [72% - 78%]) [CS40DB], 511/674 (76% [72% - 79%] ) [CS2DB], 461/674 (68% [65% - 72%]) [S2DB] and 321/674 (48% [48% - 51%]) [CDB]).

**Discussion**

We developed a robust algorithm capable of generating realistic diffusion-weighted positive synthetic stroke lesions on diffusion-weighted magnetic resonance images, and produced a large database of three-dimensional synthetic cases. We demonstrated that the segmentation performance of a generic 3D U-Net trained on a combination of human-labelled clinical stroke images and the synthetic stroke images significantly outperformed the models trained on the human-labelled clinical stroke images alone, and rivaled human abilities in terms of sensitivity. The model trained on the clinical database + 40,000 synthetic strokes (CS40DB) proved more sensitive but less specific relative to human readers and might therefore substantially assist the radiologist to avoid missing lesions. This work demonstrates that the shortcoming of having relatively few medical images for training can be overcome by producing realistic synthetic images. The method presented should be generalizable to other pathologies and could substantially improve machine learning results in medical imaging applications. Recently, data augmentation with generative adversarial networks based synthetic images have been shown to be promising for example for the training of lung nodules (16) or brain tumor detection (17), liver lesion classification (18), or abdominal organ segmentation (19) algorithms.

The Dice coefficient obtained with a standard generic 3D U-Net trained on the clinical stroke database was similar to the published networks to our knowledge, from which the most successful have a problem-specific architecture: Chen et al (20) obtained a Dice of 0.67 using their MUSCLE network, Liu et al (21) using an auto-encoder architecture based on ResNet (22) obtained a Dice of 0.645, and Kamnitsas et al (23) won the ISLES Challenges 2015 (7) with their DeepMedic Network by achieving a Dice of 0.57. In a study with a large and multicentric dataset of 2,770 patients and using the DeepMedic framework, Wu et al (24) reported median Dice coefficients of 0.79 to 0.86. They did not report the mean Dice obtained, but compared their results to be similar to the mean Dice of 0.67 obtained by Chen et al (20). Zhang et al (25) obtained a Dice of 0.79 on their validation set, but only 0.58 on the ISLES challenge test set.

Interestingly, a clear increase in network performance was seen when the number of additional synthetic volumes was increased from 2,000 to 40,000. Because the synthetic lesions were produced randomly out of a pool of 2,027 normal volumes and 375 volumes with stroke, this shows that a combinatoric effect took place, and that the method presented permits a meaningful extension to the base set on which the training was performed. Further work should investigate whether the diversity of the synthetic stroke lesions could be additionally enhanced using generative methods such as generative adversarial network (26) or variational auto-encoder (27), and if the final performance on segmentation and lesion detection could be further increased.

The Dice coefficient of the model trained on the CS40DB was close to the human inter-reader Dice of 0.76. It is important to note that the choice of the contrast parameters by human reader is a confounding factor (**Suppl. Fig.**

2). This makes the production of reference standard labels non-trivial and limits the potential for Dice improvement. A major advantage of the synthetic generation method as proposed here is the generation of training cases with perfect labels (we used here an increase of signal of at least 8% compared to background), and this should increase the robustness and consistency of the training by the network compared to lesions drawn based on subjective evaluation. Indeed, the addition of synthetic images did not only improve the Dice coefficient, but also significantly improved the stability of the results.

This work has several limitations. It was a single center study. There was a significant difference between the age of subjects in the stroke and the normal database. Some of the improvements determined by the models compared to the humans could be accounted for by the tiredness and time limitations of the human readers, but this is one of the expected strengths of machine learning derived detection and segmentation. A large number of parameters were selected by relying upon experience and good engineering principles without direct evaluation, but the goal of this paper was to compare the effect of the synthetic data generation, and not the network parameters themselves. Further, several technical challenges intrinsic to deep learning that are particular to medical images, such as noisy images with heterogeneous signal compared to standard optical images acquired with a camera, were not directly assessed here. Finally, DW hyperintense lesions might not be completely equivalent to ischemic stroke lesions (28), and many institutions assess initial acute stroke patients with CT and not MR.

In conclusion, we present a method to generate an arbitrarily large number of diffusion-weighted positive synthetic stroke lesions with perfect labels, which improves the performance of a generic 3D U-Net in detecting acute stroke lesions on diffusion-weighted images. The method presented is likely generalizable to other pathologies and could substantially improve machine learning results in medical imaging applications.

**Acknowledgments**

Christian Federau is supported by the Swiss National Science Foundation. This work was supported by the Swissheart Foundation and a SPARK Grant of the Swiss National Science Foundation. Calculation for this project were performed in part using support from Google and on a Titan Xp donated by the NVIDIA Corporation.

**References**

1.  Tsai JP, Mlynash M, Christensen S, et al. Time From Imaging to Endovascular Reperfusion Predicts Outcome in Acute Stroke. Stroke. 2018;49(4):952–957.

2.  Saver JL, Goyal M, van der Lugt A, et al. Time to Treatment With Endovascular Thrombectomy and Outcomes From Ischemic Stroke: A Meta-analysis. JAMA. 2016;316(12):1279–1289.

3.  Esteva A, Kuprel B, Novoa RA, et al. Dermatologist-level classification of skin cancer with deep neural networks. Nature. 2017;542(7639):115–118.

4.  De Fauw J, Ledsam JR, Romera-Paredes B, et al. Clinically applicable deep learning for diagnosis and referral in retinal disease. Nat Med. 2018;24(9):1342–1350.

5.  Titano JJ, Badgeley M, Schefflein J, et al. Automated deep-neural-network surveillance of cranial images for acute neurologic events. Nat Med. 2018;24(9):1337–1341.

6.  Hainc N, Federau C, Stieltjes B, Blatow M, Bink A, Stippich C. The Bright, Artificial Intelligence-Augmented Future of Neuroimaging Reading. Front Neurol. 2017;8:489.

7.  Maier O, Menze BH, von der Gablentz J, et al. ISLES 2015 - A public evaluation benchmark for ischemic stroke lesion segmentation from multispectral MRI. Medical Image Analysis. 2017;35:250–269.

8.  http://www.image-net.org. .

9.  Ronneberger O, Fischer P, Brox T. U-Net: Convolutional Networks for Biomedical Image Segmentation. arXiv:150504597 [cs]. 2015;http://arxiv.org/abs/1505.04597. Accessed September 12, 2018.

10. http://www.bic.mni.mcgill.ca/ServicesAtlases. .

11. http://picsl.upenn.edu/software/ants. .

12. https://horosproject.org. .

13. https://www.tensorflow.org. .

14. Kingma DP, Ba J. Adam: A Method for Stochastic Optimization. arXiv:14126980 [cs]. 2014;http://arxiv.org/abs/1412.6980. Accessed July 4, 2019.

15. Benjamini Y, Yekutieli D. The Control of the False Discovery Rate in Multiple Testing under Dependency. :24.

16. Han C, Kitamura Y, Kudo A, et al. Synthesizing Diverse Lung Nodules Wherever Massively: 3D Multi-Conditional GAN-Based CT Image Augmentation for Object Detection. 2019 International Conference on 3D Vision (3DV). 2019. p. 729–737.

17. Han C, Rundo L, Araki R, et al. Combining Noise-to-Image and Image-to-Image GANs: Brain MR Image Augmentation for Tumor Detection. arXiv:190513456 [cs, eess]. 2019;http://arxiv.org/abs/1905.13456. Accessed February 23, 2020.

18. Frid-Adar M, Diamant I, Klang E, Amitai M, Goldberger J, Greenspan H. GAN-based Synthetic Medical Image Augmentation for increased CNN Performance in Liver Lesion Classification. Neurocomputing. 2018;321:321–331.

19. Sandfort V, Yan K, Pickhardt PJ, Summers RM. Data augmentation using generative adversarial networks (CycleGAN) to improve generalizability in CT segmentation tasks. Scientific Reports. 2019;9(1):1–9.

20. Chen L, Bentley P, Rueckert D. Fully automatic acute ischemic lesion segmentation in DWI using convolutional neural networks. Neuroimage Clin. 2017;15:633–643.

21. Liu Z, Cao C, Ding S, Han T, Wu H, Liu S. Towards Clinical Diagnosis: Automated Stroke Lesion Segmentation on Multimodal MR Image Using Convolutional Neural Network. arXiv:180305848 [cs]. 2018;http://arxiv.org/abs/1803.05848. Accessed September 12, 2018.

22. He K, Zhang X, Ren S, Sun J. Deep Residual Learning for Image Recognition. arXiv:151203385 [cs]. 2015;http://arxiv.org/abs/1512.03385. Accessed September 14, 2018.

23. Kamnitsas K, Ledig C, Newcombe VFJ, et al. Efficient multi-scale 3D CNN with fully connected CRF for accurate brain lesion segmentation. Medical Image Analysis. 2017;36:61–78.

24. Wu O, Winzeck S, Giese A-K, et al. Big Data Approaches to Phenotyping Acute Ischemic Stroke Using Automated Lesion Segmentation of Multi-Center Magnetic Resonance Imaging Data. Stroke. 2019;50(7):1734–1741.

25. Zhang R, Zhao L, Lou W, et al. Automatic Segmentation of Acute Ischemic Stroke From DWI Using 3-D Fully Convolutional DenseNets. IEEE Transactions on Medical Imaging. 2018;37(9):2149–2160.

26. Goodfellow IJ, Pouget-Abadie J, Mirza M, et al. Generative Adversarial Networks. arXiv:14062661 [cs, stat]. 2014;http://arxiv.org/abs/1406.2661. Accessed September 12, 2018.

27. Kingma DP, Welling M. Auto-Encoding Variational Bayes. arXiv:13126114 [cs, stat]. 2013;http://arxiv.org/abs/1312.6114. Accessed September 16, 2019.

28. Inoue M, Mlynash M, Christensen S, et al. Early Diffusion-Weighted Imaging Reversal After Endovascular Reperfusion Is Typically Transient in Patients Imaged 3 to 6 Hours After Onset. Stroke. 2014;45(4):1024–1028.

Table 1. Overview of Dice Coefficient of the Test Set for the Models Trained with the Clinical and Synthetic Databases

|  | Mean ± Standard Deviation Dice | Median Dice | Min – Max Range Dice | Mean ± Standard Deviation Error in Stroke Volume [mL] | p-value versus S2DB | p-value versus CS2DB | p-value versus CS40DB |
|---|---|---|---|---|---|---|---|
| CDB | 0.618 ± 0.077 | 0.639 | 0.005 - 0.670 | 5 ± 15 | $2 \cdot 10^{-8}$ | $2 \cdot 10^{-30}$ | $2 \cdot 10^{-42}$ |
| S2DB | 0.653 ± 0.036 | 0.661 | 0.338 - 0.680 | -2 ± 3 | - | $5 \cdot 10^{-32}$ | $7 \cdot 10^{-62}$ |
| CS2DB | 0.695 ± 0.027 | 0.700 | 0.457 - 0.720 | 1 ± 1 | - | - | $2 \cdot 10^{-18}$ |
| CS40DB | 0.721 ± 0.010 | 0.725 | 0.677 - 0.743 | 1 ± 0 | - | - | - |

Note.—Dice coefficient of the test set (n=192) achieved in the 'trained range' for each model. Mean ± standard deviation, median, minimum and maximum range of the Dice score and as well as the error in stroke volume mean ± standard deviation, of the trained range from epoch 250 to 450, except for the CS40DB from epochs 100 to 300. P-value represent the paired, two-tailed, Student's t-test of the Dice coefficient of the respective model in the trained range.

Table 2. Detailed Stroke Lesion Detection Analysis

|  | Epoch | Sensitivity | Specificity | PPV | NPV |
|---|---|---|---|---|---|
| CDB | 423 | 85% [82%-87%] | 48% [44%-51%] | 64% [61%-67%] | 74% [69%-78%] |
| S2DB | 296 | 77% [73%-79%] | 68% [65%-72%] | 73% [70%-76%] | 72% [69%-76%] |
| CS2DB | 274 | 80% [77%-83%] | 76% [72%-79%] | 79% [76%-82%] | 78% [74%-81%] |
| CS40DB | 154 | 91% [89%-93%] | 75% [72%-78%] | 80% [77%-83%] | 88% [85%-91%] |
| Human reader 1 | - | 78% [75%-81%] | 92% [90%-94%] | 92% [89%-94%] | 79% [76%-82%] |
| Human reader 2 | - | 79% [76%-82%] | 89% [86%-91%] | 89% [86%-91%] | 79% [76%-82%] |
| Human reader 3 (reference standard) | - | 84% [81%-87%] | 96% [94%-98%] | 96% [94%-97%] | 85% [82%-87%] |

Note.— Detailed stroke lesion detection analysis in a random subset (n=80 patients) of the test set, with confidence interval.

| Parameters | Normal Database | Stroke Database |
| --- | --- | --- |
| Repetition Time (ms) | 6990 ± 1519 [3550 – 9600] | 7256 ± 1266 [2500 – 12000] |
| Echo Time (ms) | 100 ± 3 [59 – 104] | 89 ± 16 [57 – 120] |
| Acquisition Matrix (range) | 116 x 116 – 192 x 192 | 116 x 116 – 200 x 200 |
| Slice Thickness (mm) | 3 ± 0.2 [3 – 5] | 3.2 ± 0.6 [2 – 5] |
| Pixel Bandwidth (Hz) | 1173 ± 123 [765 – 1715] | 1169 ± 277 [590 – 1955] |
| Field Strength 3 T / 1.5 T (% of cases) | 36% / 74% | 76% / 34% |

Suppl. Table 1. MRI acquisition parameters. Values are mean ± standard deviation [range] unless otherwise specified.

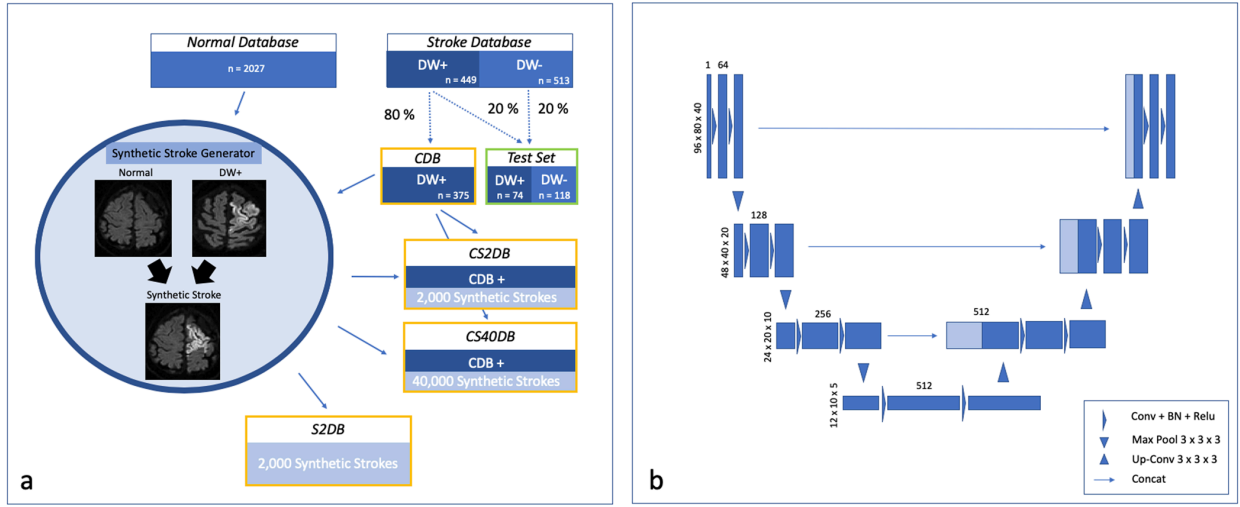

**Fig. 1. (a)** Schematic flow diagram of the databases. Boxes in dark blue indicate clinical images, while images in light blue indicate synthetic images. The boxes with the orange border represent the four different datasets on which the model was trained. The box with the green border represents the final test set. Only DW positive cases (n = 375) were included in the training set. The test set (n = 192) included both DW positive (n = 74) and negative (n = 118) cases. Note that because of the difference in underlying anatomy, the lesion in the synthetic stroke images has a different appearance to the lesion in the clinical image. **(b)** Schema of the architecture of the 3D U-Net used with a depth of three layers. The first layer had 64 features maps; this doubled in each consecutive layer. Blue rectangles represent four-dimensional tensors (3D image volume with size indicated on the left x 1-dimensional number of features indicated on top).

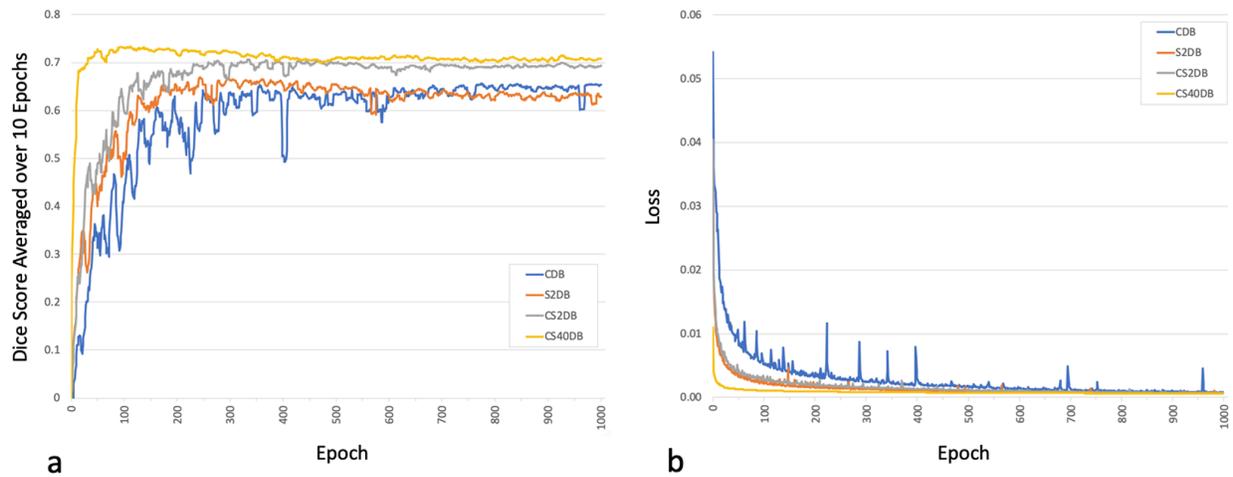

**Fig. 2. (a)** Dice score on the test set averaged over 10 epochs, for the four models trained on the four different databases. The model trained on the CS40DB (yellow curve) visibly outperformed all other models. Note also the significant increase in model stability when trained on the CS40DB, and to a lesser extent when trained on the CS2DB. The model trained on the S2DB outperformed the model trained on the CDB in the early phase of training, before overfitting occurred. The corresponding non-averaged curves can be found in **Suppl. Fig 5**. **(b)** Training loss (weighted cross-entropy + Dice Loss) as a function of number of training epochs.

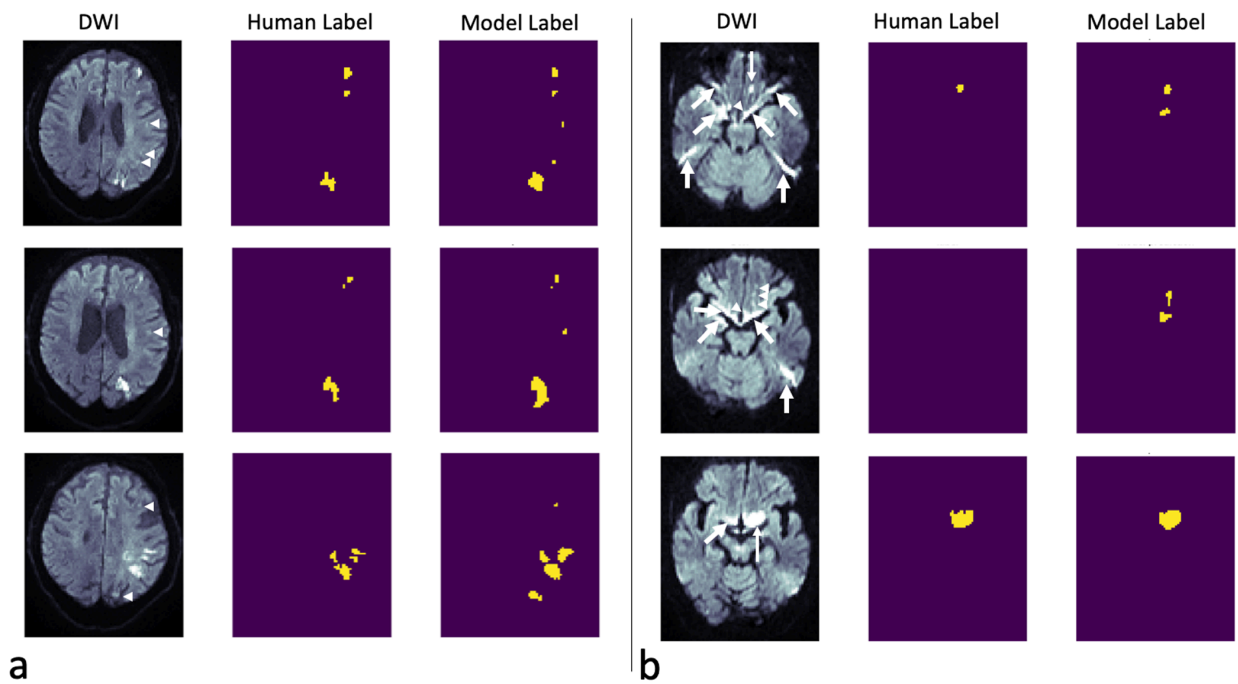

**Fig. 3.** Examples of human labelling (middle columns) to the model trained on the CS40DB (right columns). **(a)** The model marked several lesions missed by the human reader (arrowheads). **(b)** Both the human reader and the model marked a lesion in the left gyrus rectus (long arrow, top row), and in the region of the left hypothalamus (long arrow, bottom row). In addition, the network also marked several discrete lesions (arrow heads, middle and top rows), that it correctly did not confound with susceptibility artefacts (thick arrows).

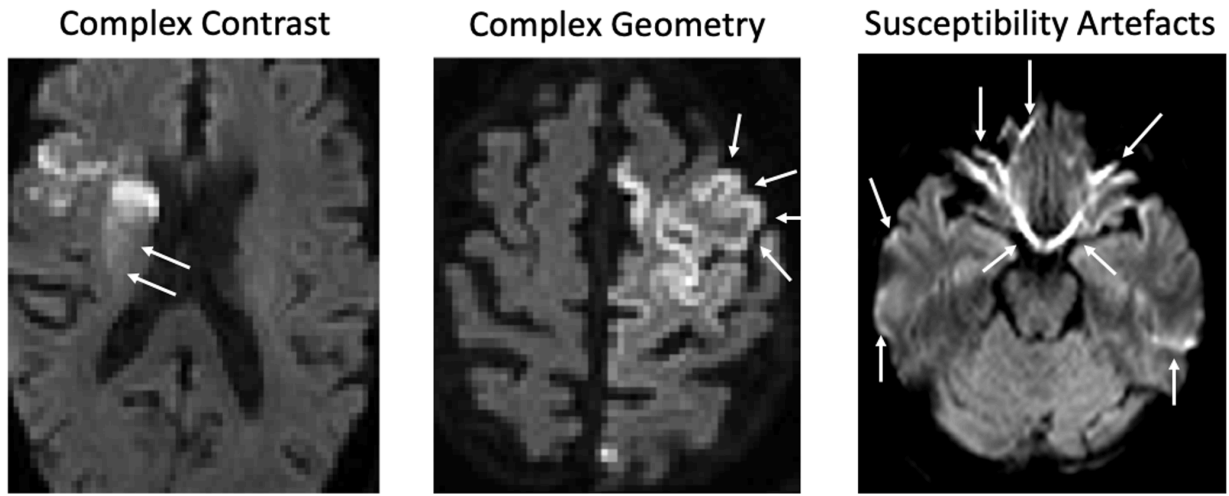

**Suppl. Fig. 1**. Some of the challenges in segmenting diffusion-weighted stroke lesions include complex contrast with unclear borders (left), complex lesion geometry (center) and susceptibility artefacts that have similar contrast to stroke lesions (right).

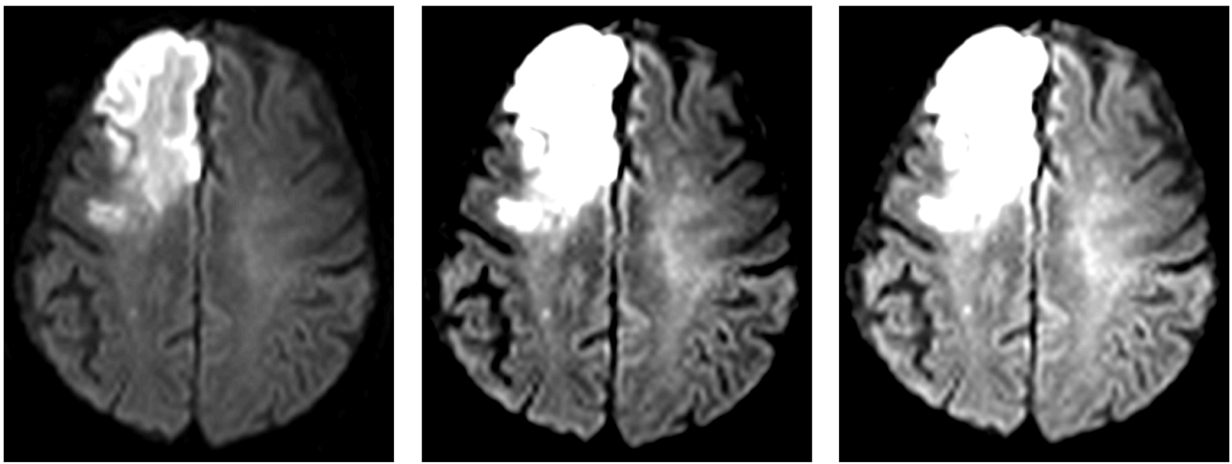

**Suppl. Fig. 2**. Diffusion-weighted image of the same stroke lesion of the right frontal lobe shown with three different contrast adjustments, which is typically adjusted subjectively by the neuroradiologist in the daily clinical routine, demonstrating that the boundaries of the lesion are dependent on the subjective contrast.

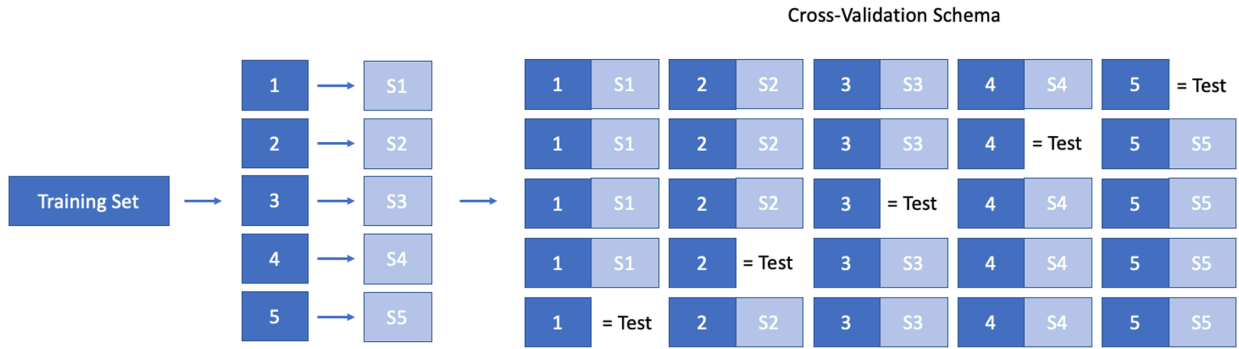

**Suppl. Fig. 3**. Cross-validation: The training set of the clinical database was divided in 5 equal buckets of 75 cases (dark blue, 1-5). In each bucket, 75 corresponding additional synthetic cases (light blue, S1-S5) were produced out of the clinical cases of each bucket. The cross-validation training of the 3D U-Net was then performed on 4 buckets comprising a total of 600 cases (i.e. with an equal amount of clinical and synthetic cases). The testing was performed on the clinical cases only of the bucket left over. This was cycled through the 5 possible bucket combinations, and results were averaged.

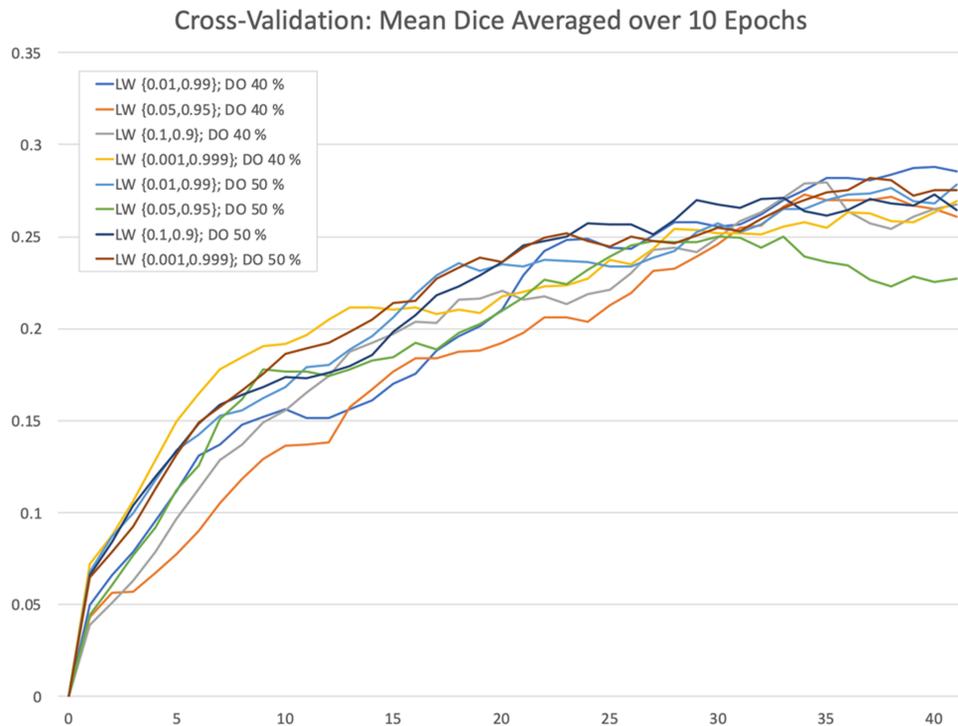

**Suppl. Fig. 4**. Cross-validation mean Dice averaged over 10 epochs. The curves look similar for almost all of the parameter sets. The parameter set loss weights of {0.01, 0.99} and dropout of 40% were selected for the further analysis. LW = Loss weights. DO = Dropout.

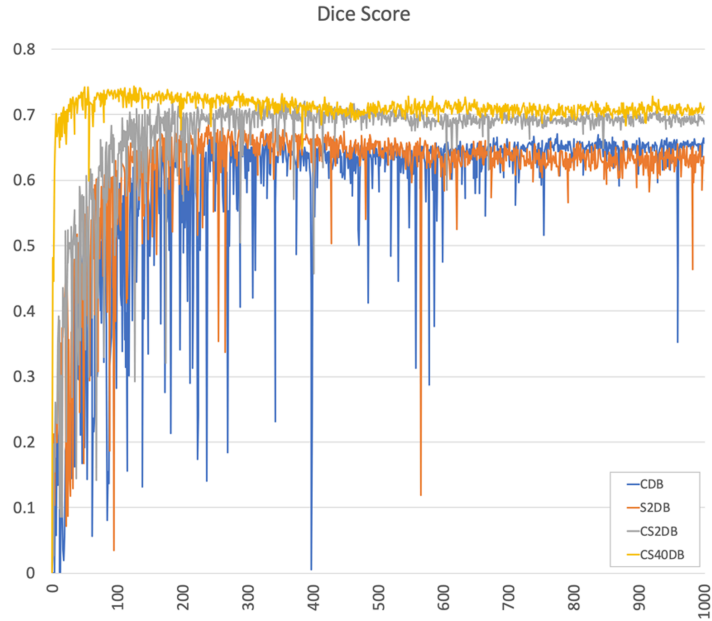

**Suppl. Fig. 5**. Dice score of the test set as function of number of training epochs.

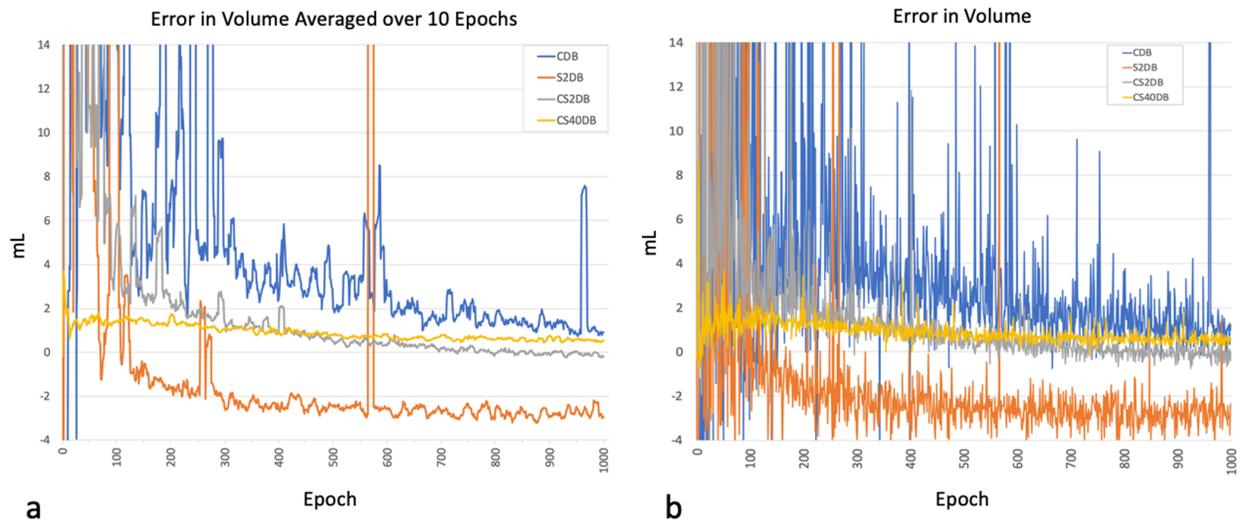

**Suppl. Fig. 6**. Volume error of the test set as function of number of training epochs **(a)** averaged over 10 epochs and **(b)** for each epoch. Note that the model trained on the S2DB systematically slightly underestimated the stroke volume compared to expert opinion, while the error in volume slowly asymptomatically diminished for the models trained on the other databases.

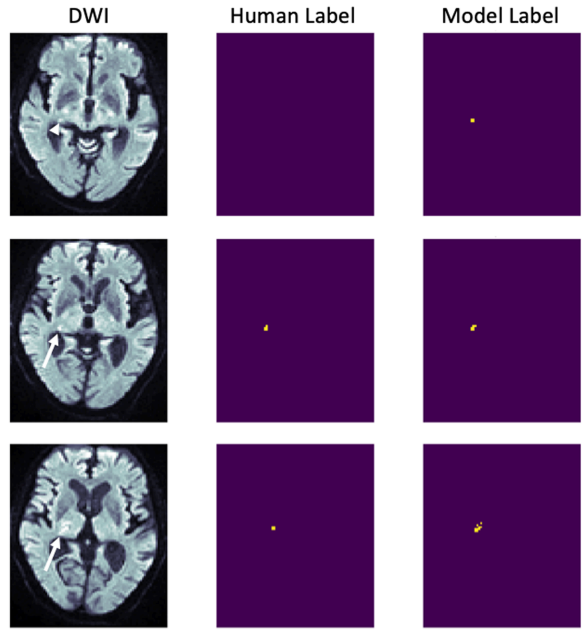

**Suppl. Fig. 7**. A very small lesion (arrow head, top row) was marked by the model trained on the CS40DB but not by the human experts. The lesion was better visible on adjacent slices (long arrow, middle and bottom row).

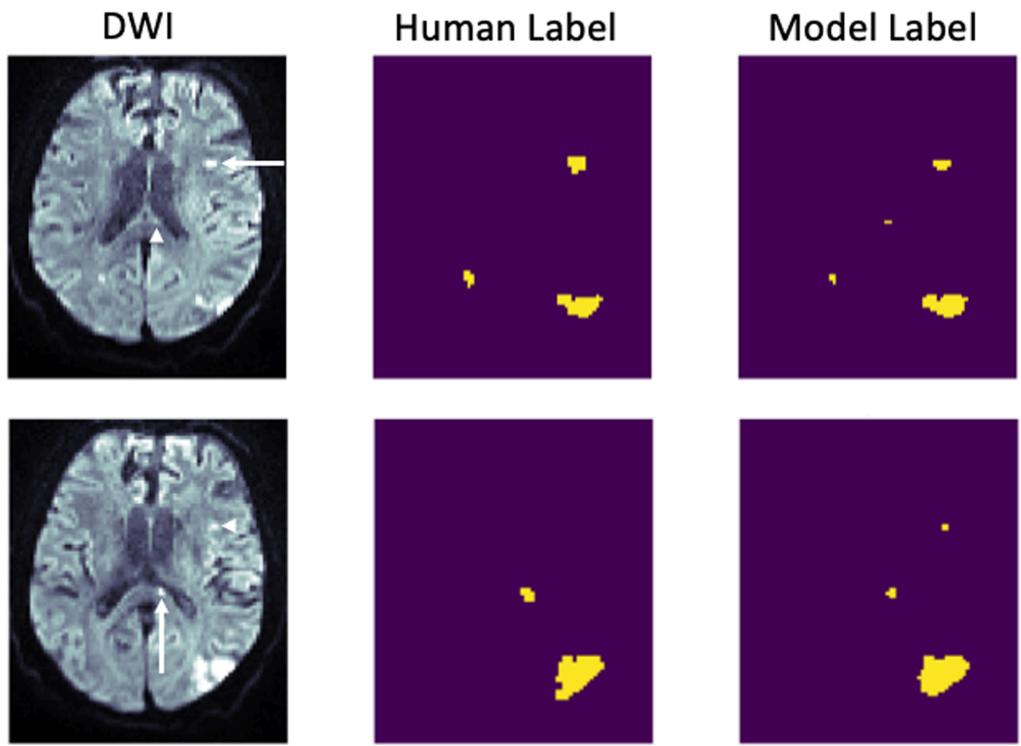

**Suppl. Fig. 8**. Small lesions of the corpus callosum (arrow head, top row) and the anterior part of insula (arrow head, bottom row) were missed by the human experts but marked by the model trained on the CS40DB. Both lesions were better visible on adjacent slices (low arrows).